\documentclass[prb,a4paper,twocolumn,showpacs,showkeys,superscriptaddress]
{revtex4}
\usepackage{amsfonts, amssymb, amsmath, graphicx, psfrag}
\begin{document}
\newcommand{\eq}[1]{\begin{equation} #1 \end{equation}}
\newcommand{\gi}[1]{$\Phi_{\text{#1}}$}
\newcommand{\m}[1]{$\mu_{\text{#1}}$}
\newcommand{\E}[1]{$E_{\text{#1}}$}
\newcommand{\fig}[1]{\begin{figure}[htb!] #1\end{figure}}
\title{Interrelation of work function and surface stability: the case of
 BaAl$_4$}
\date{\today}
\author{M.~A.~Uijttewaal}
\email{M.Uijttewaal@science.ru.nl}
\affiliation{Radboud University, IMM, P.O.~Box 9010, 6500 GL, Nijmegen, The Netherlands }
\author{G.~A.~de~Wijs}  
\affiliation{Radboud University, IMM, P.O.~Box 9010, 6500 GL, Nijmegen, The Netherlands }
\author{R.~A.~de~Groot}
\affiliation{Radboud University, IMM, P.O.~Box 9010, 6500 GL, Nijmegen, The Netherlands }
\affiliation{Univ Groningen, MSC, Chem Phys Lab, Nijenborgh 4, Groningen, NL-9747 AG, The Netherlands}
\author{R.~Coehoorn} 
\affiliation{Philips Research Laboratories, Prof.~Holstlaan 4, 5656 AA,
 Eindhoven, The Netherlands}
\author{V.~van~Elsbergen} 
\author{C.~H.~L.~Weijtens} 
\affiliation{Philips Research Laboratories, Weisshausstr 2, D-52066 Aachen,
 Germany}
\begin{abstract}
The relationship between the work function ($\Phi$) and the surface stability 
of compounds is, to our knowledge, unknown, but very important for applications such as organic light-emitting diodes. 
This relation is studied using first-principles calculations on various surfaces of BaAl$_4$.  
The most stable surface \mbox{[Ba terminated (001)]} has the lowest $\Phi$ (1.95~eV), which is lower than that of any elemental metal including Ba. 
Adding barium to this surface neither increases its stability nor lowers its work function. 
BaAl$_4$ is also strongly bound. 
These results run counter to the common perception that stability and a low $\Phi$ are incompatible. 
Furthermore, a large anisotropy and a stable low-work-function surface are predicted for intermetallic compounds with polar surfaces. 
\end{abstract}

%\centerline{\textbf{Table of Contents Synopsis}}
%~\\

%The relationship between the work function ($\Phi$) and the surface stability of compounds, which is very important for electron-emitting applications, is studied for the first time using first-principles calculations on surfaces of BaAl$_4$. From the results, stable low-work-function surfaces are predicted for intermetallic compounds with polar surfaces. \\

%\vspace{2cm}

%\centerline{\includegraphics[angle=-90, width=9.4cm]{Figure3.jpg}}

%\newpage

\maketitle

\centerline{\textbf{I. Introduction}}
%~\\

The long-standing problem of the precise relationship between the work function
($\Phi$) and the stability of metals has become pressing with the increased
application of electron-emitting materials in technology. Two examples are 
vacuum electronic devices like cathode-ray tubes (CRTs) and cathodes for organic
light-emitting diodes (OLEDs). In CRTs, a thin layer of a low-$\Phi$ metal is
 often present on top of a cathode made from a structurally stable material 
to enhance its electron-emitting properties.\cite{CRT} Electron injection into 
OLEDs strongly depends on the cathode work function,\cite{Camp} while the 
lifetime of the device can be limited by the stability of the cathode material. 
OLEDs with an alloy interface between the cathode and the polymer have been 
found to be superior, in terms of lifetime and luminosity, to those with 
single-element cathodes.\cite{pLED, pLED2} Thus the relationship between $\Phi$
and stability\cite{stab} is crucial. However, it is poorly understood, 
especially for more complex metals.%\\
%~\\
  
\fig{\centerline{\framebox{\includegraphics[angle=0, width=7.5cm]{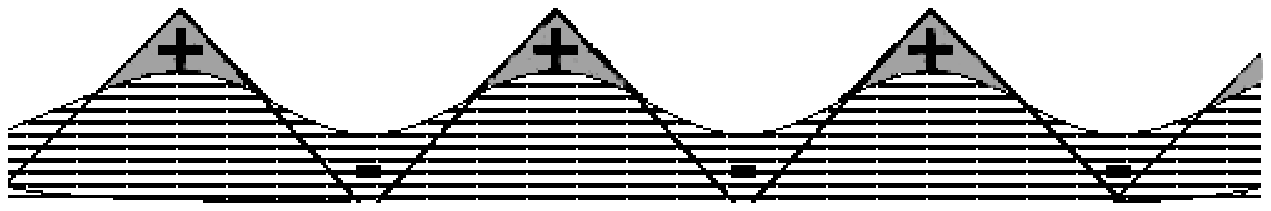}}}
\caption{\label{smooth}Schematic illustration of Smoluchowski smoothing of the 
electron charge density (striped area) at an open surface. The grey area is 
formed by bulk Wigner-Seitz cells. Signs indicate the net charge. The dipole 
thus formed lowers the work function.}
}

The general rule for elements is that a low work function and high stability are
incompatible: the element with the lowest work function, cesium ($\Phi$ = 
2.14~eV),\cite{WFa} is highly reactive and has a low melting temperature. Noble
metals (silver/gold/platinum) on the other hand are hardest to oxidize, but 
their $\Phi$ is at least twice as large (4.25/5.1/5.65~eV).\cite{WFa} 
It is generally believed that this must always be the case, \emph{i.e.}, that a low 
$\Phi$ implies loosely bound electrons that easily mediate reactions.
The work function surface anisotropy, however, can be quite large: for tungsten,
it is of the order of 1~eV.\cite{WFW}
The anisotropy was already theoretically addressed by Smoluchowski.\cite{Smol} 
According to his model, at a ``more open''\cite{open} surface, relaxation of 
the electrons ``smoothes'' the surface charge.  A dipole moment is built up that
lowers $\Phi$.\cite{dipole} This is schematically shown in Fig.~\ref{smooth}. 
Although the work function is lowered for a more-open surface, the surface 
energy increases and stable, low-work-function surfaces are forbidden by the 
model. Hardly any extensions to the model have been suggested since,\cite{fAlL}
certainly not for more complex metals. Experimental results for transition metal
alloys suggest the so-called alloy effect,\cite{alloy} which implies that the 
stability and $\Phi$ of alloys interpolate between those of the constituting 
elements. As a consequence, it is believed that stable, low-work-function 
surfaces are not possible for more complex metals either.

Nevertheless, the work function and stability of compounds mainly constitute 
a \emph{terra incognita}. In this paper, the relationship between $\Phi$ and 
stability of compounds is studied for the first time by calculating \emph{ab 
initio} the structural relaxation, work function and surface energy of various 
surfaces of BaAl$_4$. The motivation to study the BaAl$_4$ system was the 
successful use of Ba-Al-alloy cathodes in OLEDs and the high melting point of this compound ($>1000^\circ$C). In the Ba-Al phase diagram,\cite{Hans} the melting point of 
BaAl\textbf{\emph{$_4$}} is the highest, much higher than those of the 
constituents, which indicates a strongly bound structure. Moreover, we show 
that the work function for one of the crystal surfaces is very low.\\
%~\\

\centerline{\textbf{II. Ab Initio Calculations}}
%~\\

 The BaAl$_4$ crystal structure\cite{struc} is depicted in Fig.~\ref{BaAl4}. It
 is body-centered tetragonal, with alternately three aluminum layers (1Al, 2Al 
and 3Al) and one barium layer (1Ba) in the [001] direction. Both Al and Ba atoms
are located approximately according to their elemental crystal structures (bcc 
and fcc, respectively). The (100) surface is stoichiometric and there are four 
(001) surfaces. They are constructed by cutting the bulk \emph{above} the 
accordingly labeled layers in Fig.~\ref{BaAl4}. Two other (001) surfaces are 
also considered: ``2Ba'' (one bcc barium layer added to 1Ba) and 
``$\frac{1}{2}$Ba'' (half a barium layer removed from 1Ba).%\\

 \fig{\begin{minipage}[l]{0.35\linewidth}
\includegraphics[width = 5 cm, angle = 0]{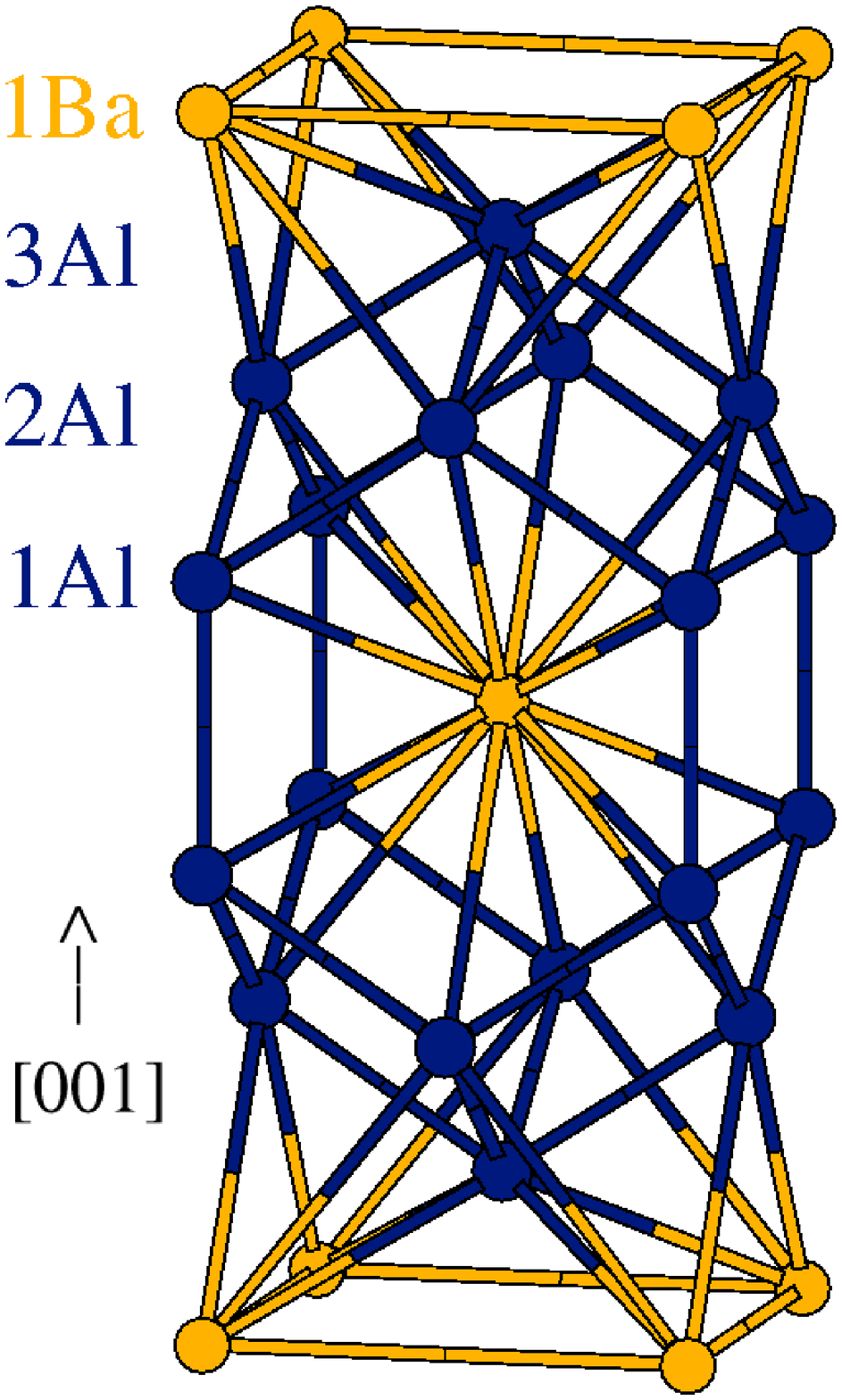}
\end{minipage}
\hspace{1cm}
\begin{minipage}[r]{0.35\linewidth}
\caption{\label{BaAl4}Body centered tetragonal unit cell of 
BaAl$_4$ containing 2 formula units. The labels refer to the layer and atom 
type. 1Al and 3Al atoms are equivalent in the bulk.}
\end{minipage}
}

%~\\
From bulk calculations we find that $a =  4.56$~\AA , that $c = 11.39$~\AA\ , and 
that the remaining free parameter in the structure, the height of the third 
aluminum layer, is $0.381\cdot c$. These values compare very well (deviations 
$\approx 1$\%) with those obtained experimentally.\cite{struc}  
The bulk density of states (DOS) shows a quasi-gap just below the Fermi level,
 in accordance with the previously calculated DOS of BaAl$_4$ and other 
 isoelectronic compounds with the same structure.\cite{struc}  
The valence electrons are mainly located on Al atoms. The binding energy 
(\E{bind}) is 1.42~eV/F.U.\ with respect to the elemental bulk metals. Together with the quasi-gap, it hints at BaAl$_4$'s stability.

The first-principles calculations were carried out using density functional 
theory (DFT) in the local density approximation (LDA)\cite{DFT,DFT2} with 
generalized gradient corrections (GGA).\cite{GGA}
We used the total energy and molecular dynamics program called VASP (Vienna 
\emph{Ab-initio} Simulation Package),\cite{VASP,VASP2} which has the 
projector-augmented-wave method (PAW)\cite{PAW,PAW2} implemented.
Nonlinear core corrections\cite{core} were applied for both barium and 
aluminum. A semi-core of Ba $5s$ and $5p$ electrons was included. The Kohn-Sham 
orbitals were expanded in plane waves with cutoffs of 18~Ry. The Brillouin zones
for the (100) and (001) surface calculations were sampled with 
$1\times16\times8$ and $12\times12\times1$ Monkhorst-Pack\cite{Monk} $k$ point 
grids, respectively, resulting in eight and 21 $k$ points, respectively, in 
their irreducible parts. The (periodically repeated) unit cell for the (100) 
surface calculation contained a slab with a thickness of seven bulk unit cells 
and 16 \AA\ vacuum. The supercells for the (001) surface calculations contained
slabs with thicknesses of six formula units (F.U.) and 20 \AA\ vacuum. \\  %~\\

\centerline{\textbf{III. Work Function}}

\fig{\centerline{\includegraphics[angle=-90, width=8.4cm]{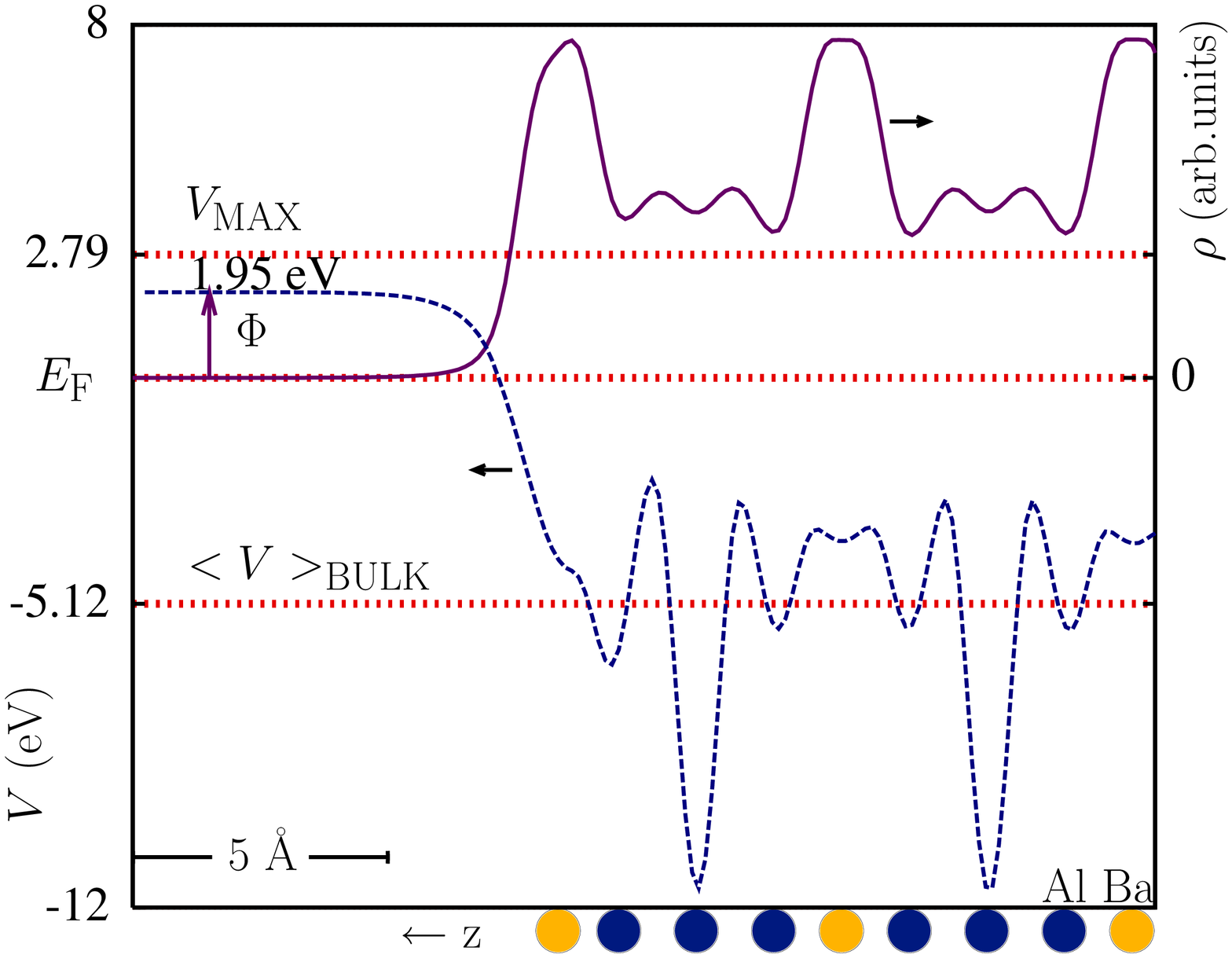}}
\caption{\label{PC001BaAl}The BaAl$_4$ (001) 1Ba surface. 
Laterally averaged charge density ($\rho$ solid line, arbitrary units) and 
(electrostatic) potential ($V$ dashed line, eV relative to $E_{\text{F}}$) as 
function of position perpendicular to the surface ($z$). The positions of Ba and Al layers are
indicated with light and dark spheres, respectively. At $-5.12$~eV is the average 
bulk potential and at 2.79~eV is the highest potential in the bulk.
\protect{\cite{ref}} The work function is 1.95~eV.}
}

%~\\
The work function is defined as the amount of energy it takes to extract 
electrons from a metal, \emph{i.e.}, bring them from the Fermi level to the vacuum. 
At locations that are microscopically far from the material, but macroscopically
near it, $\Phi$ is surface dependent. The work function at large distance is 
then an average over the various surfaces.\cite{Fall} 
The work function of a surface is calculated by constructing a supercell with a 
slab of material, with only this surface, and vacuum. Fig.~\ref{PC001BaAl}  
illustrates this using the (001) Ba terminated surface. About 10~\AA\ empty space 
suffices for the electrostatic potential\cite{XC} to converge to its 
vacuum value ($V_{\text{vac}}$). A thick slab, however, is required for the 
Fermi level (\E{F}) to be accurate. A more efficient method is to link the 
surface calculation to a bulk one where \E{F} is very accurate. The average 
potential in the bulk ($<V>_\text{bulk}$) is set equal to the average potential 
in the middle of the slab. Accuracies of a few hundredths of an eV can thus be 
achieved with only six to eight bulk unit cells, depending on direction.
\cite{acc} The work function of the 1Ba surface is 1.95~eV.

\fig{\centerline{\includegraphics[angle=-90,width=8.4cm]{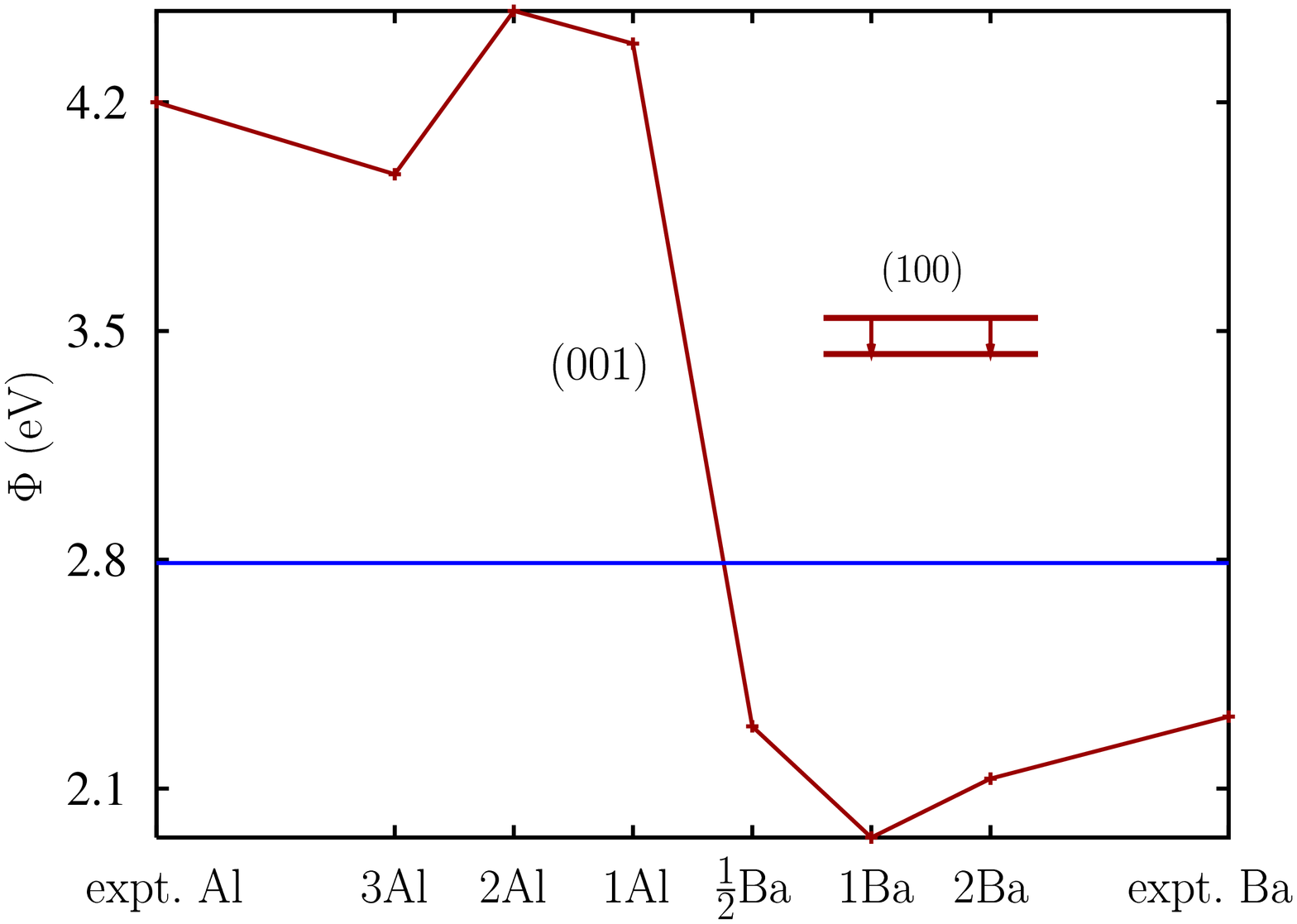}}
\caption{\label{WFBa}Work functions (eV) of the BaAl$_4$ (001) 
surfaces. Experimental bulk values for barium and aluminum are indicated at the 
borders. Lines connect the data points. $\Phi$ for the (100) surface before and 
after relaxation is inserted. (The dotted line is $V_\text{max}$)\protect{\cite{ref}}}
}

The work functions of the other BaAl$_4$ surfaces are calculated analogously. 
That of the (100) surface is reduced by structural relaxation from 3.54~eV to 
3.43~eV. The work functions of the (001) 1Al, 2Al and 3Al surfaces are 4.38~eV,
4.48~eV and 3.98~eV, respectively. Relaxation has negligible effect.
The work functions of the various surfaces are summarized in Fig.~\ref{WFBa}. 
As plotted there, covering the Al surface with half a monolayer of barium (thus
forming the $\frac{1}{2}$Ba surface) already greatly reduces the work function
(2.29~eV). The 1Ba work function is reduced even more to a surprisingly low 
value of 1.95~eV. This is below the $\Phi$ of any element.\cite{WFa} 
An additional layer of barium (thus forming the 2Ba surface) increases $\Phi$ 
(2.13~eV) again. Experimental bulk values for Al (4.2~eV) and Ba (2.32~eV) are 
plotted for reference.\cite{WFa} These compare favorably with calculated work
functions of bcc-barium [2.36~eV (100) and 2.27~eV (111)] and fcc-aluminum 
[4.34~eV (100) and 4.17~eV (111)].

The huge ($>2$~eV) variation in the (001) surface work function can be 
(qualitatively) explained from the difference in the atomic electronegativities.
Ba is less electronegative (0.9)\cite{elneg} than Al (1.5),\cite{elneg} 
effecting charge transfer from barium to aluminum. The resulting surface dipole
moment decreases the work function of the Ba surface to even under the 
elemental barium bulk value and increases $\Phi$ of the Al surfaces 
considerably. The dependence of $\Phi$ on Al coverage of the 1Ba surface can be 
understood if one realizes that the 1Al and 3Al surfaces are more open than the 
2Al surface. According to the Smoluchowski model, $\Phi$ must then be lowered. 
The decrease in $\Phi$ during relaxation of the (100) surface can also be understood by
combining charge transfer with Smoluchowski smoothing. The barium atoms at the 
surface move out ($\approx$~0.1~\AA ), as they favor an environment with a 
lower charge density, while the aluminum atoms at the surface tend to get 
closer together, as they favor an environment of a high charge density. We 
conclude that a mono-layer coverage yields an extreme work function.%\\~\\

\centerline{\textbf{IV. Surface Energy}}
%~\\

The (relative) stability of a surface is determined by the difference in surface
energy ($\gamma$) between two surfaces.\cite{stab} The surface energy
is calculated as the energy of a slab with only this surface minus the energy of
the bulk equivalent (formed by merging the [periodically repeated] slabs 
together), divided by two times the surface area, because a slab has two sides. 
This method was used for the (100) and \mbox{(001) $\frac{1}{2}$Ba} surfaces. 
Slabs for the other (001) surfaces are non-stoichiometric and so there is no 
equivalent bulk. Calculating $\gamma$ now requires reservoirs of Ba and Al.
Assuming thermodynamic equilibrium, the chemical potential of the aluminum 
(\m{Al}) and that of the barium (\m{Ba}) reservoir are linked to the total 
energy per F.U.\ of BaAl$_4$ (\E{bulk}): 
\eq{\label{eq1}E_{\text{bulk}}=4\cdot\mu_{\text{Al}}+\mu_{\text{Ba}}.}
The energy of, for example, the 1Ba surface, as a function of \m{Ba}, follows from
the total energy of a slab with these surfaces exclusively ($E_{\text{1Ba}}$):
\cite{North} 
\begin{equation}
\gamma^{\text{1Ba}}(\mu_{\text{Ba}})=[E_{\text{1Ba}}-\text{\#F.U.}\cdot 
E_{\text{bulk}} - \mu_{\text{Ba}}]/2A_{\text{S},}
\end{equation}
where $A_{\text{S}}$ is the surface area.%\\~\\

\fig{\centerline{\includegraphics[angle=-90,width=8.4cm]{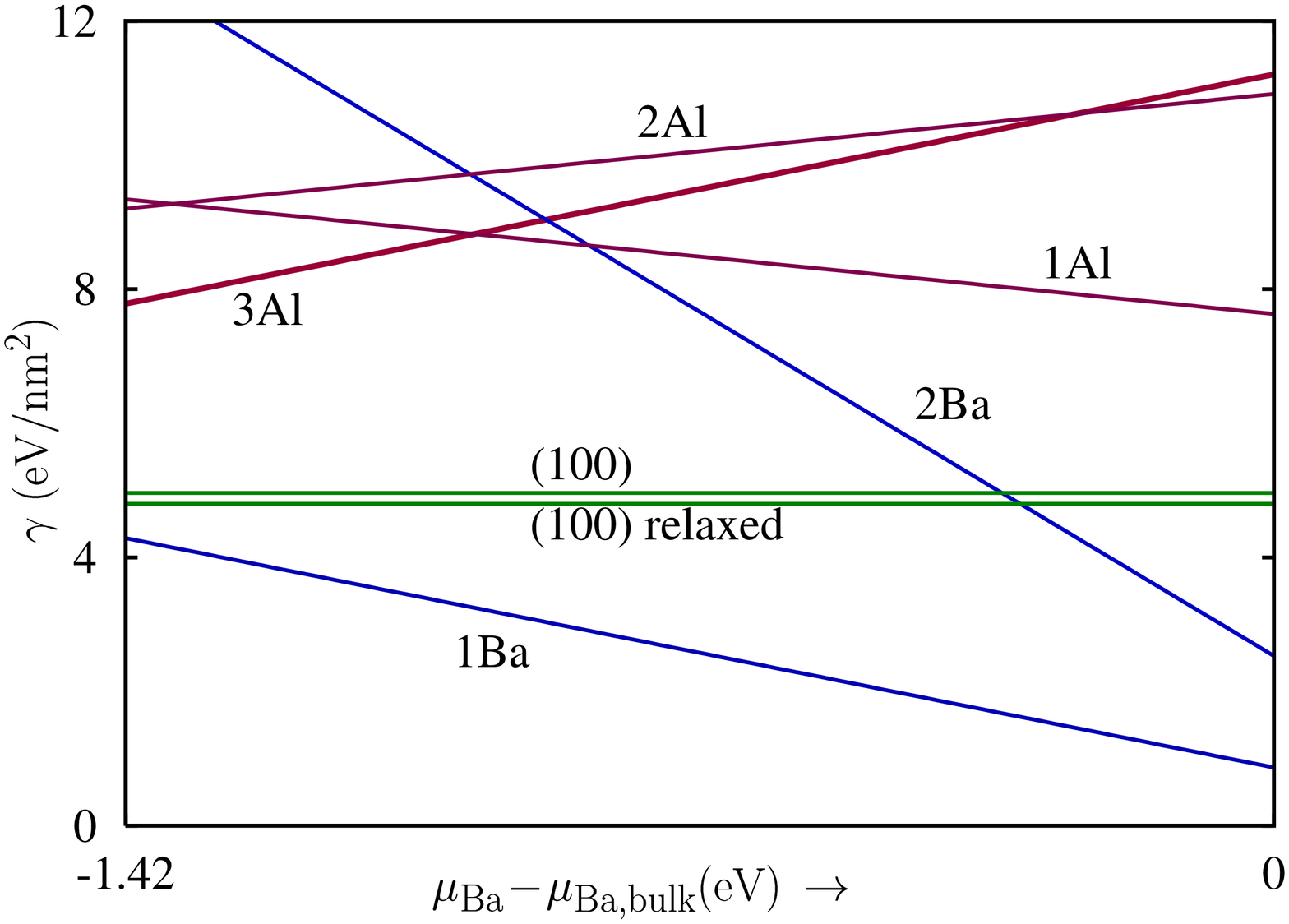}}
\caption{\label{Es001}Surface energies (eV/nm$^2$) for the 
BaAl$_4$ surfaces as function of the barium chemical potential (eV). The 
potential ranges from the value for Al bulk to that of Ba bulk. The difference 
is the binding energy of BaAl$_4$ (equation~\ref{eq1} has been used). For the 
(100) surface the effect of relaxation is indicated. The (001) 1Ba surface is 
the most stable.}
}

%~\\
The surface energies are drawn in Fig.~\ref{Es001}. Structural relaxation of the
(100) surface lowers its energy from 4.96~eV/nm$^2$ to 4.82~eV/nm$^2$. The 
energy of the (001) $\frac{1}{2}$Ba surface equals 13.3~eV/nm$^2$ and is outside
the range of the plot. For the other (001) surface, $\gamma$ depends on the 
barium chemical potential. In the figure, $\mu_\text{Ba}$ varies over the 
thermodynamically allowed range between the Ba and Al bulk chemical potentials
\cite{vdWal} as phase separation occurs outside this range. The single
barium surface is the most stable one in the entire region and so other (001) surfaces cannot be formed. 

The unusual stability of the low-$\Phi$ surface is explained in the same way as 
the relaxation of the (100) surface. Since barium is the less electronegative element, it favors a low electron density and makes the most stable surface. 
Additional barium layers at the surface decrease this stability, as Ba-Ba bonds 
are less strong than Al-Ba bonds, especially when the barium in contact with 
aluminum is (partially) ionized.\\
%~\\

\centerline{\textbf{V. Conclusions.}}
%~\\

As both the low $\Phi$ of the 1Ba surface and its stability followed from Ba's 
lower electronegativity, we come to a remarkable prediction: For an 
intermetallic compound with polar surfaces, like BaAl$_4$, the most stable surface has the 
lowest work function and relaxation can only decrease it further.

To summarize, we used first-principles calculations on various surfaces of 
BaAl$_4$ to study the interrelation of work function and stability. The
 bulk work function of BaAl$_4$ is 2.79~eV,\cite{ref} its anisotropy (1.5~eV) is huge, and the most stable surface (Ba terminated 001) has the lowest $\Phi$ (1.95~eV), which is even lower than that of pure Ba. Adding barium to this surface neither increases its stability nor lowers its work function. The binding energy of 1.42~eV/F.U., the quasi-gap in the DOS, and its melting temperature of over 1000 $^\circ$C indicate BaAl$_4$'s stability.
These results contradict the common perception that stability and a low $\Phi$ are incompatible. 
They also run counter to the effect that alloy work functions are in between those of the constituting elements. 
Its stable, very-low-$\Phi$ surface and the stable structure probably make BaAl$_4$ a good electron-emitting material. 
Furthermore, a stable low-work-function surface promises to be general for intermetallic compounds with polar surfaces. 
\begin{acknowledgments}%\textbf{Acknowledgments.}
We would like to thank A. van Dijken and E. Meulenkamp 
for valuable comments. This work was part of the research program of the 
Stichting voor Fundamenteel Onderzoek der Materie (FOM) with financial support 
from the Nederlandse Organisatie voor Wetenschappelijk Onderzoek (NWO).
\end{acknowledgments}

%\newpage


\begin{thebibliography}{30}
\bibitem{open}An open surface is understood as one with a relatively small 
number of atoms per unit area.
\bibitem{dipole}The energy of a surface dipole is the work it costs to move an 
electron through this dipole. Natural as it may seem, the charge-smoothing 
dipole moment cannot be used to split the work function into a bulk and a 
surface part, as is often done in literature. Within the bulk, the potential
is periodic (no dipole) and outside it there is no charge. It follows that the 
reference for this dipole is the average (electrostatic) potential on the outer 
surface of the bulk unit cells, which means that it is surface-dependent. It is
discussed in this paper.
\bibitem{XC}The exchange correlation part of the potential is not needed because
it vanishes in the vacuum. Furthermore, including it is impractical as its 
range is longer.
\end{thebibliography}
\end{document}